# Manual and Fast C Code Optimization

**Mohammed Fadle Abdulla**
**Computer Science and Engineering Department,**
**Faculty of Engineering, University of Aden, Aden, Yemen**
al_badwi@hotmail.com

**ABSTRACT:** Developing an application with high performance through the code optimization places a greater responsibility on the programmers. While most of the existing compilers attempt to automatically optimize the program code, manual techniques remain the predominant method for performing optimization. Deciding where to try to optimize code is difficult, especially for large complex applications. For manual optimization, the programmers can use his experiences in writing the code, and then he can use a software profiler in order to collect and analyze the performance data from the code. In this work, we have gathered the most experiences which can be applied to improve the style of writing programs in C language as well as we present an implementation of the manual optimization of the codes using the Intel VTune profiler. The paper includes two case studies to illustrate our optimization on the Heap Sort and Factorial functions.

**Introduction**

Software applications are designed to work. Hence, the program functionality is the mean objective of the programmer during the phase of development of a new application. The performance consideration may add to the design later on.

There are many techniques in which a given software program can be made to run faster. However, speeding up a program can also cause an increase in the program size. In the world of mobile systems, the graphics library should be small enough to run on the mobile device without compromising graphics quality. Devices like mobiles, PDAs etc, have strict memory restrictions [MR98].

So, the mean objective of the performance improving is to write the code in such a way that memory and speed both be optimize. Several different options exist for measuring application performance. Homegrown timing functions inserted into the code are a more effective way to gather performance data. Other most efficient and accurate ways to gather timing data is to use a good performance profiler, which show the time spent in each function of the program and will also provide an analyses based on this data [***05, Mye99].

Once performance data has been collected, it needs to be analyzed. Find the routines that are taking the majority of the application time, and the loop that appears to be taking more time than it should. These areas are known as "Hotspots". There are a lot of tools available for detecting the hotspots [AN04], and the optimization should be done on these hotspots.

The next step is to test the modified code, if still run correctly with the achieved performance improvement. Otherwise, the optimization cycle is repeated for another iteration.

In this work, we have gathered the most experiences which can be applied to make a C code optimized for speed as well as memory. We have also used Intel VTune profiler to find out where program spends most time. The paper includes two case studies to illustrate our optimization on the Heap Sort and Factorial functions.

**1. Basic performance events**

The VTune profiler monitors all the programs executing on the system including the operating system, JIT-compiled Java applications, .NET applications, and device drivers. Sampling does not modify your binary files or executables in order to monitor the performance of your application [M+06].

Here, in this paper, we select some of the basic events which occur during the execution of an application, to be monitored for the application performance evaluation, namely, the Core Cycles (CPU_CLK), Instructions Retired (INST_Retired), and Cycles per Instruction (CPI) [Chu98, Bud94].

Instructions Retired (INST_Retired):
This event indicates the number of instructions that retired or executed completely. This does not include partially processed instructions. The Sampling Interval is the frequency of data collection. For *Time-based* *sampling*, the Sampling interval represents the amount of wall-clock time the Analyzer waits before collecting each sample. For *Event-based sampling (EBS),* the Sampling interval is used to calculate the target number of samples and the Sample_After_value. The following formula is used to calculate the average number of samples collected per second per event.

$$Number\ of\ Samples = \frac{Duration(\sec) \times Number\ of\ Processors}{Sampling\ Interval\ (\sec)}$$

This number of samples collected per second is multiplied by *Duration* to get the target number of samples that are collected for an event in an Activity run. The *Sample_After_value* is calculated based on the target number of samples for an event.

Cycles per Instruction (CPI):
As a general guide these numbers have been derived from experienced performance engineers: from 1 to 5. A high value for this ratio indicates that over the current code region, instructions are taking a high number of processor clocks to execute. This could indicate a problem. A high CPI value still indicates a performance problem however a high CPI value on a specific logical processor could indicate poor CPU usage and not an execution problem.

During sampling, a sample (instruction address) is collected periodically for each event until the specified duration is complete. Types of information collected for a particular event in this sampling session are:

### *Event_name samples:*
Event data in samples are provided by default in all sampling views.

### *Event_name %*
This column contains the percentage of samples collected for the event. This value is calculated using the following formula:

$$Event\_name\ \% = \frac{Number\ of\ samples\ collected\ for\ the\ event}{Total\ number\ of\ samples\ collected} \times 100$$

95

*Event_ name Events:*
The number of times the event occurred during sampling data collection. The sampling collector determines this value by applying the following formula for each event:

$$\boxed{\begin{array}{l} \text{Event\_ name } E\text{vents} = \\ \text{Event\_name Samples} \times \text{Sample\_After\_Value} \end{array}}$$

The total event count calculated using this formula provides an estimate of the total number of times an event occurred.

**2. Manual performance improving**

In this work, we illustrate two ways to improve the performance of an application. First, we spot and rewrite those parts of the program which are usually consuming large number of clock ticks (samples). Secondly, the VTune profiler is used for father locating the slowest parts within the program (hotspots), which will help us to improve them.

By locating and improving separated parts of the program will help in improving the overall performance of the application. In the following, we illustrate the most programming practices which can make the C program to run fast, namely:

- **Variable data types:**

Most of the processors can handle *unsigned* integer arithmetic considerably faster than *signed* one. Therefore, it is a good programming practice to use *unsigned int* instead of *int* if we know the value will never be negative. Also, integer arithmetic is much faster than floating-point arithmetic, as it can usually be done directly by the processor, rather than on external coprocessor or math libraries.

- **Register Variable allocation:**

Global variables are never allocated to registers. Global variables can be changed by assigning them indirectly using a pointer, or by a function call. Hence, the compiler cannot cache the value of a global variable in a register, resulting in extra (often unnecessary) loads and stores when globals are used. We should therefore not use global variables inside critical loops. If a function uses global variables heavily, it is beneficial to copy those global variables into local variables so that they can be assigned to registers. This is possible only if those global variables are not used by any of the functions which are called.

- **Loop Termination Condition:**

The loop termination condition can cause significant overhead. It more efficient to write *count_down_to_zero* loops and uses simple termination conditions Take the following two loops; the first uses an incrementing loop, the second a decrementing loop.

*for ( i = 1 ; i <= n ; i++) { …}*
*for ( i= n ; i != 0 ; i-- ) { …}*

The second one will be faster. Or we could also write the loop as

*For ( i= n ; i-- ; ) { …}*

which is the same, but more efficient because it is quicker to process i (as the test condition). For the original code, the processor has to calculate

*subtract i from n, Is the result non_zero? So, increment i and continue..."*

- **Branch Removal and Function Inlining:**

Functions always have a certain performance overhead when they are called. The program counter has to be changed, the arguments variables have to be pushed onto a stack, and new variables allocated. For example, consider the following function:

*for ( i = n/2 ; I >= 1 ; i--)*
 *adjust( a, i, n) ;*
                *…*
*void adjust( TYPE *a, int I ,int n) { …….}*

Function ***adjust()*** is often called from a loop, it is better to put that loop inside the function to cut down the overhead of calling this function repeatedly. This function can be written as:

*adjust ( a) ;*
                *…*
   *void adjust( TYPE *a)*
   *{*
   *for ( i = n/2 ; i >= 1 ; i--) { …}*
             *}*

- **AND/OR Conversion:**

Most of the processors can handle *bitwise operations* considerably faster than *logical* one. It often suggests substituting bitwise AND "&" and OR "|" for

97

logical AND "&&" and OR "||" in the C program.

In general, savings can be made by trading off memory for speed. If you can cache any often used data rather than recalculating or reloading it, it will help.

- ❖ Minimize the use of global variables, and declare anything within a file (external to functions) as static, unless it is intended to be global.
- ❖ Use word-size variables if you can, as the machine can work with these better instead of char, short, double, bit fields etc.
- ❖ Don't use recursion. Recursion can be very elegant and neat, but creates many more function calls which can become a large overhead.
- ❖ Single dimension arrays are faster than multi-dimension arrays.
- ❖ Use #defined macros instead of commonly used tiny functions

## 3. Evaluation

This section illustrates the steps of implementation of the optimization technique on two selected algorithms, namely, the Heap sort and the factorial function. At the beginning the profiler runs to create a simulation session. A simulation session contains the clock-tick database, the relative instruction frequencies. the function-call table and information regarding address and execution interlocks. All this information is presented graphically and one can get more information about an interlock or the issues related to a particular instruction by clicking on it. .The Code Coach is a nice feature coupled with V-Tune. It takes the preprocessed source file and gives useful hints to optimize the code at the C code level.

During dynamic analysis, a particular part of the program has been run many times to see the various cache and memory access issues related with the execution. But this is a very slow process because after each instruction, the dynamic analyzer intervenes to check the state of the registers and buffers.

The process of performance improving is started by writing the code for the algorithm and then compiles and debugs it. We then generate a program database file that contains the symbol information VTune needs. The *Time_Based_Sampling* gives us the hotspots in the code. A hotspot indicates sections of code within a module that had highest activity during the Monitor Session. Activity implies the number of samples collected and the amount of CPU time spent executing that section of the module.

When the session is over we get the module report for Clock ticks window, which shows the CPU image and Hotspots in module window where the hotspots of our application are displayed. We are interested in finding the function where the program spends most of its CPU time (Using Hotspots by function option), within the function. We can also find the Hotspots_By_Location to locate the parts to be optimized. We also used the dynamic analysis to identify the instructions that cause critical performance problems. A FOR loop in our code in selected to run the simulation with the Dynamic_Code_Analysis Setup option[8].

### 3.1. Case Study I: Optimizing the Heap Sort Function

As an example, we have taken a sample application that performs the heap sorting. The program sorts 100,000 random integers. In the Code sample 1 (following), we provide the source code for the application.

**Code Sample 1: The source code of the Heap sort program (HeapS$_0$)**

```
#include "header"
#define TYPE  int     /* type of key by which to sort */
#define N 100000      /* number of elements to sort */
/* prototypes of functions to be used later */
void gen_array(TYPE **a, int n);
void hsort(TYPE *a, int n);
void adjust(TYPE *a, int i, int n);
void swap(int *a, int *b) ;
int  main()    /* main function */
{
TYPE *a;
#ifdef  DEBUG
int i;
#endif
gen_array(&a, N);    /* generate random numbers as raw data */
#ifdef DEBUG
for (i=l; i <= N; i++)      printf("%d ", a[i]);printf("\n");
#endif
hsort(a, N); /* sort the array */
#ifdef DEBUG
for (i=l; i <= N; i++)      printf("%d ", a[i]);
printf("\n");
#endif
return 1;
}
```

```
void  gen_array(TYPE **a, int n){
int i;
*a = (int *)  malloc((n+1) * sizeof(TYPE)); /* allocate memory */
srand(time((time_t *)NULL));  /* initialise the random num generator */
for (i=1; i < n; i++)     (*a)[i] = rand();    /* generate the numbers */
(*a)[0] = n;
}
void  hsort(TYPE *a, int n)  {     /* n = number of elements */
int i;    TYPE t;
for (i=n/2; i >= 1; i--)   /* form the heap */
adjust(a, i, n);
/* now we have a heap */
   for (i=n-1; i >= 1; i--)/* repeat */
     {
swap(&a[1], &a[i+1]); /* swap a[i+1] and a[l] */
adjust(a, 1, i); /* recalculate heap */
} }
void swap(int *a, int *b) {    int   t;   t = *a; *a =*b; *b = t;    }
void adjust(TYPE *a, int i, int n) {
/* adjust a root node in left and right heaps */
int  j,   done = 0;    TYPE k;   /* n = total elements, i = root node to adjust */
k = a[i]; j= 2*i;
while ( (j <= n) && !done)  {
if (j < n)     if (a [j] < a[j+1])       j = j+1; }
if (k >= a[j])       done = 1;
else  {   a[j/2] = a[j] ;    j = 2*j;}}
a[j/2] = k;
} /* of adjust */
```

This example illustrates bad programming style! It contains lots of errors that have negative impact on the performance.  Profiling will help to find them all.

The performance measure before optimizations was 579.3 samples. The updated program under the profiler shows that the number of hotspots has been decreased to 500.  The manual optimization is performed as follows:

*Function inlining:*

The ***swap()*** function in this program has been inlined manually. Function inlining reduces the number of function calls, and is especially useful in a loop. The performance gain by this was the reduction in the number of

Anale. Seria Informatică. Vol. VIII fasc. I - 2010
Annals. Computer Science Series. 8<sup>th</sup> Tome 1<sup>st</sup> Fasc. – 2010

samples by 21. This can be done either in macro definition in C or by manual inlining of the C functions. In this work we applied the inling keyword of C code.

*Register variable allocation:*

In a statement like *if ( ( j < n) & (a[j] < a[j+l]) )* the value of **j** is loaded thrice. But if **j** is explicitly declared to be a register variable, three loads get reduced to one. This optimization resulted in a reduction of 52 samples.

*Branch removal:*

Various branches can be combined, as is done at the beginning of the ***adjust()*** function. Since this is happening inside a loop, it significantly reduces the time spent inside the loop. This resulted in a total reduction of 35 samples

*AND/OR conversion:*

Logical AND/OR can be converted to bitwise AND/OR, if possible. Bitwise AND/OR is supposedly faster than logical AND/OR. In this case, it did not affect the performance much.

*Loop Termination Condition:*

Converting the termination condition to count_down_to_zero has remarkable improvement. This technique results in a reduction of 52 samples.

    The optimized version gave a reading of 465.3-52 samples. The optimized version is given below.

```
   int main()    /* main function */
   { …
     for (i=N; i != 0; i--)
          printf("%d ", a[N-i+1]);
    …
   for (i= N; i !=0 ; i--)
          printf("%d ", a[N-i+1]);
   printf("\n") ;
   #endif
   return 1;  }

   void gen_array(register TYPE **a, int n)
    { …
     for (i= n-1; i !=0 ; i--)
   (*a)[n-i] = rand();            /* generate the numbers */
   (*a)[0] = n;    /* 1st element  = num of elements */
    }
```

```
void  hsort(register TYPE *a, int n)    {
 /* n = number of elements */
register int i;  TYPE t ;
 …
          }
void  adjust(register TYPE *a, register int i, int n)
   {   register int j; int done = 0; TYPE k;
    …
   } /* of adjust */
```

## 3.2. Case Study II: Optimizing the Factorial Function

The second program selected for the manual optimization is the Factorial function. The program calculates the factorial of 2000. The source code sample 2 of the program is given below.

**Code Sample 2:  The source code of the Heap sort program**

```
#include    "header"
#define FA_SIZE 10000
 /* number of array elements */
#define FA_DPE 4
  /* 4 digits per array element */

int fa [FA_SIZE];
int fa_modulo; /* 10 - FA_DPE */
 int count;
void fact(int n);
void print_fa( );
void mult_fa(int n);
void init_fa();

int  main(int argc, char **argv) {
   fact(2000);
  /* Calculate factorial of 2000 */
   return 0;   }

void fact(int n) {
    int i;
```

```
    fa_modulo = 1;
for (i=1; i <= FA_DPE; i++)        fa_modulo *= 10;
for (i=2; i<=n; i++)      mult_fa(i);
        #ifdef  DEBUG
            print_fa()
        #endif   }
void mult_fa(int k)
{
register int  i = 0;   int carry = 0;   int product = 0;
do {
product = fa [i] * k + carry;
fa[i] =  product % fa_modulo ;
carry = product / fa_modulo;
 i++;
}
while (i <= count  || carry> 0)   count = i-1;
}
void init_fa(){
int i;
for (i=1; i<FA_SIZE; i++)    fa[i]  = 0;
  fa[0] = 1;
}
void print_fa ()
 {
char str[10] = "" ;  int i;
printf("%0d", fa[count--]);
  while (count>=0)
{
sprintf (str, "%0d",  fa[count]);
for (i=FA_DPE; i > strlen(str); i--)
putchar('0') ;
printf( "%0d", fa[count]);
  fflush(stdout) ;
count--;
}
printf("\n") ;
}
```

The performance before optimization was 709.2 samples. The optimizations performed were as follows:

*Register Variable Allocation:*
In this program, we have various global variables which are frequently used but cannot be declared register variables. So, in a function which uses these variables, we create a register alias for such a variable. This resulted in a reduction of samples to 668.5.

*Using optimized assembly routines:*
The *initfa()* routine uses a loop to initialize the fa array. Instead, we can use an optimized assembly function like *memset* to do the job. This resulted in a small reduction of 22 samples (to 646.5 samples).

*Branch removal:*
This had almost no effect in this case.

*Logical AND  JOR conversion:*
This also had a very small effect of just 10 samples

*Global Variables:*
Since global variables are never allocated to registers and they can be changed only indirectly using a pointer, or a function call, extra loads and stores are required when they are used. In *multi_fa(..)* function, we had copy *Count* variable into local variable *LocalCount*, which can be assigned to register.

> *Register int LocalCount = Count ;*

Therefore instead the global *Count* value be loaded and stored each time it is incremented, the LocalCount variable in the modified version, needed only a single instruction. This result in a total reduction of 45 samples.

*Loop Termination Condition:*
Converting the termination condition to count_down_to_zero has remarkable improvement. This technique results in a reduction of 38 samples. The optimized version thus performed at 636-38-45 samples. The optimized version is as under:

```
int main (int argc, char **argv)
    {
  fact(2000);    /* Calculate factorial of 1000 */
  return 0;         }
void fact(int n) {
   register int i;    fa_modulo = 1;
      for (i= FA_DPE; i !=0 ; i--)  fa_modulo *= 10;
    ...
```

```
    }

void mult_fa(int k)   {
      register int i=0;
int carry = 0;
int product = 0;
register int c = count;
register int LocalCount = count;
register int *faa = fa;
do
{  ...
 ...}
while ((i <= c) | (carry> 0))    LocalCount = i-1;
count = LocalCount;
    }

void init_fa()   {
 memset((char *)(fa+1), 0, FA_SIZE-1);
 fa[0] = 1;      }

void print_fa()   {
     register int LocalCount = count;
char str[10]= "";    int i;
printf("%0d", fa[LocalCount --]);
 while (LocalCount >=0)
 {   sprintf(str, "%0d", fa [LocalCount]) ;
      for (i=FA_DPE; i > strlen(str); i--)  putchar ('0') ;
     printf("%0d", fa[LocalCount]);   fflush(stdout);
              LocalCount --;   }
     printf ("\n");
     count = LocalCount;        }
```

## 4. Results

The process of the performance improvement is taken on a sample application that performs the heap sorting which sorts 100,000 random integers. The time duration is taken as 20 seconds with the Interval of 1 msec time. The calculated value of the Sample_After_Value is 2000000.

The optimization is running in three iterations with three versions of the original heap sort program (HeapS), namely, (i) the version Heap_Optimized1 where the FOR loops are optimized, (ii) the version Heap_Optimized2 where the Heap_Optimized1 is further improved with the optimization of the local and general registers allocations, and (iii) the version Heap_Optimized3 where the Heap_Optimized2 is further optimized with branch removal and Inline technique.

**Table 1: The modified program Heap_Optimized2**

|  | Module (Process Heap_Optimized2) | | | | | |
|---|---|---|---|---|---|---|
|  | Adjust | Hsort | Swap | Gen_array | Rand | memSet |
| CPU_CLK samples | 31 | 4 | 2 | 1 | 0 | 1 |
| INST_Retired samples | 27 | 0 | 4 | 0 | 1 | 0 |
| CPU_CLK % | 79.49% | 10.26 | 5.13 | 2.56 | 0 | 2.56 |
| INST_Retired.ANY % | 84.38% | 00 | 12.5 | 0 | 3.3 | 0 |
| CPU_CLK events | 62000K | 8000K | 4000K | 2000K | 2000K | 2000K |
| INST_Retired events | 54000K | 0 | 8000K | 0 | 0 | 0 |
| % of the Process | 86.11% | 6.94 | 5.56 | 0 | 1.39 | |

In the iteration of the modified program Heap_Optimized2, the analysis is for the module Adjust() is as follows:
- CPU_CLK samples:
Out of a total of 39 samples, 31 samples was collected for Adjust()
- CPU_CLK % :
79.49% out of 100% of Timer samples was collected for Adjust().
- CPU_CLK events:
The Sample After value for the Timer event is 2000000. Using the formula, Number of samples X Sample After value, the number of occurrences of the Timer event for Adjust() is 62000000.
- Process % :
86.11% out of 100% of the CPU processes consumed in executing the module Adjust(), followed by the module Swap() (which consumed 7.32%), the next positions in the list (which are consumed 2.44%) are shared by Gen_array and Rand() functions.

The performance figures for the different modules within the modified program Heap_Optimized2 is shown in Table 1. Table 2 illustrates the result for the next modification of the program, namely, Heap_Optimized3. The complete performances view of the three versions of the original HeapSort algorithm is shown in Table 3. The actual improvement in the code performances within each technique for the HeapSort and Factorial algorithms are shown in Table 4.

**Table 2: The modified program Heap_Optimized3**

|  | Module (Process Heap_Optimized3) | | | | | |
|---|---|---|---|---|---|---|
|  | Adjust() | Hsort() | Swapt () | Gen_array | Rand.c | memSet |
| CPU_CLK_Unhalted_Core samples | 35 | 2 | 2 | 1 | 1 | 1 |
| INST_Retired.ANY samples | 33 | 0 | 2 | 1 | 0 | 0 |
| CPU_CLK_Unhalted_Core % | 83.3% | 4.76 | 4.76 | 2.38 | 2.38 | 2.38 |
| INST_Retired.ANY % | 91.67% | 0 | 5.56 | 2.78 | 0 | 0 |
| CPU_CLK_Unhalted_Core events | 70000K | 4000K | 4000K | 2000K | 2000K | 2000K |
| INST_Retired.ANY events | 66000K | 0 | 4000K | 2000K | 0 | 0 |
| % of the Process | 86.37% | 2.44 | 7.32 | 2.44 | 2.44% | |

**Table 3: Performances for the original and the modified versions of HeapSort Algorithm**

|  | Process | | | |
|---|---|---|---|---|
|  | HeapSort | Heap_Optimized1 | Heap_Optimized2 | Heap_Optimized3 |
| Clockticks samples | 58 | 56 | 55 | 11 |
| Instructions Retired samples | 28 | 27 | 29 | 6 |
| Cycles per Instructions ( CPI ) | 1.242 | 1.22 | 1.19 | 0.72 |
| Clockticks % | 4.26% | 4.11 | 4.04 | 0.81% |
| Instructions Retired % | 4.83% | 4.66 | 5 | 1.03 |
| Clockticks Events | 139200K | 134400K | 132000K | 26400K |
| Instructions retired events | 67200000 | 64800000 | 69600000 | 14400000 |

**Table 4 : Improvement with different techniques**

| Algorithm | No. of Reduction in Samples | | | | | | Total Reduction (%) |
|---|---|---|---|---|---|---|---|
|  | Function InLining | Register Allocation | Branch Removal | AND/OR Conv | Loop Termin | Global Variables |  |
| Heap_sort | 22 | 40 | 44 | 3 | 22 | 22 | 12% |
| Factorial | 20 | 38 | 20 | 4 | 15 | 12 | 16% |

**Conclusions**

We have seen how various small changes in the program can dramatically affect in performance: In particular, instructions inside a loop should be very carefully written so as to minimize loop length. A tool like VTune or the Code Coach can help a lot in this process. On Unix systems, this kind of profiling is not possible because of security considerations; but still, profilers like *gprof* can give statistics regarding function call traces, time spent on each instruction and the relative frequency of instructions.

Programmer should always turn compiler optimization on. The compiler will be able to optimize at a much lower level than can be done in the source code, and perform optimizations specific to the target processor.

Using a good compiler and some knowledge about optimization techniques, developers can much more easily create high performance applications.